\documentclass[conference]{IEEEtran}

\usepackage{multirow}    
\usepackage{pifont}      
\usepackage{color}       
\usepackage{xspace}      
\usepackage[keeplastbox]{flushend}     
\usepackage{graphicx}
\usepackage{caption}
\usepackage{subcaption}
\usepackage{url}
\usepackage{rotating}
\usepackage{paralist}

\usepackage{enumitem}
\usepackage{soul}
\usepackage{algorithm}  
\usepackage{algorithmicx}  
\usepackage{algpseudocode}  
\usepackage{amsmath} 
\algnewcommand{\LineComment}[1]{\State \(\triangleright\) #1}
\usepackage{color}
\usepackage{tikz}
\usepackage{xcolor}
\usepackage[misc]{ifsym} 
\def\BibTeX{{\rm B\kern-.05em{\sc i\kern-.025em b}\kern-.08em
    T\kern-.1667em\lower.7ex\hbox{E}\kern-.125emX}}
\begin{document}

\newcommand{\circled}[1]{\tikz[baseline=(myanchor.base)] \node[circle,fill=.,inner sep=1pt] (myanchor) {\color{-.}\bfseries\footnotesize #1};}
\newcommand{\mypara}[1]{\vspace{2pt}\noindent\textbf{{#1: }}}
\newcommand{\eat}[1]{}  
\newcommand{\name}{Our Work\xspace}
\newcommand{\lowerambush}{memory ambush\xspace}
\newcommand{\upperambush}{Memory Ambush\xspace}
\newcommand{\authcomment}[3]{\textcolor{#3}{#1 says: #2}}
\newcommand{\zhi}[1]{\authcomment{zhi}{#1}{blue}}
\newcommand{\yueqiang}[1]{\authcomment{yueqiang}{#1}{red}}
\newcommand{\minghua}[1]{\authcomment{minghua}{#1}{green}}
\renewcommand{\baselinestretch}{0.96}   

\renewcommand{\algorithmicrequire}{\textbf{Input:}} 
\renewcommand{\algorithmicensure}{\textbf{Output:}} 
\newcommand{\codename}{{\emph{DRAMDig}}\xspace}

\definecolor{R}{RGB}{0,0,150}

\title{\codename: A Knowledge-assisted Tool to Uncover DRAM Address Mapping}


\author{\IEEEauthorblockN{Minghua Wang\IEEEauthorrefmark{1},
Zhi Zhang\IEEEauthorrefmark{2},   Yueqiang Cheng$^{\textrm{\Letter}}$\IEEEauthorrefmark{1} and
Surya Nepal\IEEEauthorrefmark{3}\\
 Minghua Wang and Zhi Zhang are joint first authors.}
\IEEEauthorblockA{
\IEEEauthorrefmark{1}Baidu Security, Beijing, China. 
\IEEEauthorrefmark{2}University of New South Wales, Australia and Data61, CSIRO, Sydney, Australia. \\
\IEEEauthorrefmark{3}Data61, CSIRO, Sydney, Australia. \\
\IEEEauthorrefmark{1}\{wangminghua01,chengyueqiang\}@baidu.com,
\IEEEauthorrefmark{2}zhi.zhang@data61.csiro.au,
\IEEEauthorrefmark{3}surya.nepal@data61.csiro.au. \\
}}

\maketitle


\begin{abstract}


As recently emerged rowhammer exploits require undocumented DRAM address mapping, we propose a generic knowledge-assisted tool, \codename, which takes domain knowledge into consideration to efficiently and deterministically uncover the DRAM address mappings on any Intel-based machines. We test \codename on a number of machines with different combinations of DRAM chips and microarchitectures ranging from Intel Sandy Bridge to Coffee Lake. Comparing to previous works, \codename deterministically reverse-engineered DRAM address mappings on all the test machines with only 7.8 minutes on average. Based on the uncovered mappings, we perform double-sided rowhammer tests and the results show that \codename induced significantly more bit flips than previous works, justifying the correctness of the uncovered DRAM address mappings.
\eat{ 
Rowhammer is a destructive software-induced hardware fault, which can be exploited by an unprivileged attacker to break the fundamental memory isolation schemes (e.g., user-kernel isolation). To efficiently and effectively trigger the rowhammer fault, \emph{all} existing rowhammer attacks require DRAM address mapping. As major CPU manufactures do not publicly document the mapping algorithm, reverse-engineering the DRAM address mapping is a necessity. 

In this paper, we propose a reliable knowledge-assisted tool, coined \codename, which takes DRAM information into consideration to effectively and efficiently uncover the DRAM address mapping. 
First, the tool acquires the exact number of \emph{channels, DIMMs, ranks, banks, rows, columns}
assisted by DRAM knowledge. Based on a timing channel~\cite{moscibroda2007memory}, the tool then uncovers the bits of physical addresses that index rows and columns, respectively. By applying a partition policy, the tool resolves the mapping from remaining bits to channels, DIMMs, ranks and banks. 
We test our tool on a number of machines with different combinations of DRAM chips and microarchitectures ranging from Sandy Bridge to Coffee Lake. Compared to previous works~\cite{pessl2016drama,xiao2016one}, our tool can deterministically reverse-engineer a complete DRAM address mapping on every test machine. It can quickly present the address mapping within 7.8 minutes on average (only 7.8 minutes on average).
Based on the uncovered address mappings, we perform double-sided rowhammer tests on the above machines. The experiment results indicate that our tool can induce significantly more bit flips than previous works~\cite{pessl2016drama,seaborn2015exploiting}. 
We plan to open-source our tool to help users understand the DRAM address mapping and evaluate the impact of rowhammer attacks.}

\end{abstract}


\section{Introduction}\label{sec:intro}


DRAM is the main memory unit of modern computer systems and organized into rows. In 2014, Kim et al.~\cite{kim2014flipping} reported 
a software-induced DRAM fault, the so-called ``rowhammer'', that is,  
intensive accessing two DRAM rows can cause bit flips in an adjacent row even without accessing the row. The fault was soon after exploited in cross memory boundary such as user-kernel~\cite{seaborn2015exploiting} and VM-hypervisor~\cite{xiao2016one}. To demonstrate a rowhammer exploit successfully, understanding how physical addresses are mapped to DRAM is a necessity~\cite{seaborn2015exploiting, pessl2016drama,xiao2016one} to efficiently and effectively induce rowhammer bit flips.

Although such mapping is available in AMD's architectural manual but not published by another major chip company, Intel. 
As shown in Table~\ref{tab:comparison}, Seaborn et al.~\cite{seaborn2015exploiting} was the first to uncover the mapping. They first perform a blind rowhammer test, results of which are used to analyze DRAM address mapping of a specific machine. Although their methodology is intuitive and simple, the blind test is quite inefficient (within hours) and needs to be performed again if the machine setting changes (e.g., its microarchitecture or DRAM chips are replaced). 
To address the efficiency issue, Xiao et al.~\cite{xiao2016one} applied a timing channel~\cite{moscibroda2007memory} to efficiently uncover DRAM address mapping. However, their tool only works in a limited number of machine settings (we actually experimented the code they shared and details are in Section~\ref{sec:dram mapping}).
DRAMA~\cite{pessl2016drama} was the first to present a generic reverse-engineering tool that can be used in any Intel machines. It proposes a brute-force approach to enumerate all possible combinations of physical address bits and verify each combination based on the aforementioned timing channel. 
Due to its blind selection of physical addresses, DRAMA is inefficient (within hours) and often fails to output a deterministic DRAM address mapping (we tested their code~\footnote{https://github.com/IAIK/drama} and details are in Section~\ref{sec:eva}). 

\begin{table}[!htbp]
\footnotesize
\begin{tabular}{cccc}
\hline
\multirow{2}{*}{\textbf{Uncovering Tool}} & \multirow{2}{*}{\textbf{Generic}} & \multirow{2}{*}{\textbf{Efficient}} & \multirow{2}{*}{\textbf{Deterministic}}\\  
 &  &  &    \\ \hline
\multirow{2}{*}{Seaborn et al.~\cite{seaborn2015exploiting}} & \multirow{2}{*}{${\times}$} & \multirow{2}{*}{${\times}$ (within hours)} & \multirow{2}{*}{${\surd}$} \\
 &  &  &   \\ \hline
\multirow{2}{*}{Xiao et al.~\cite{xiao2016one}} & \multirow{2}{*}{${\times}$} & \multirow{2}{*}{${\surd}$ (within minutes)} &  \multirow{2}{*}{${\surd}$} \\
 &  &  &   \\ \hline
\multirow{2}{*}{DRAMA~\cite{pessl2016drama}} & \multirow{2}{*}{{${\surd}$}} & \multirow{2}{*}{${\times}$ (within hours)} & \multirow{2}{*}{${\times}$}  \\
 &  &  &   \\ \hline
\multirow{2}{*}{\textbf{\codename}} & \multirow{2}{*}{\textbf{${\surd}$}} &  \multirow{2}{*}{\textbf{{${\surd}$ (within minutes)}}} & \multirow{2}{*}{\textbf{${\surd}$}} \\
 &  &  &   \\ \hline
\end{tabular}
\caption{A comparison of uncovering tools. \codename is the first generic and efficient tool to generate deterministic DRAM address mappings. All other tools cannot achieve these three properties at the same time.}
\label{tab:comparison}
\end{table}

\mypara{Our contributions}
In this paper, we revisit the limitations and advantages of the aforementioned algorithms and observe that none of them made full use of domain knowledge. 
With this key observation, we propose a generic knowledge-assisted tool, which utilizes domain knowledge (see Section~\ref{sec:domain}) to produce a deterministic DRAM address mapping on different machines. 
First, \codename detects physical-address bits that index rows and columns, respectively. Second, the tool resolves the mappings from remaining bits to banks including channels, DIMMs, and ranks. Third, \codename determines the bank bits that are also row bits or column bits.

We test \codename on 9 different machine settings and it is able to efficiently and deterministically uncover the DRAM address mapping on each setting within only 7.8 minutes on average.
Based on the uncovered mappings, we perform double-sided rowhammer tests on three different machine settings and all results show that \codename is much more effective in inducing rowhammer bit flips than the other generic tool, DRAMA~\cite{pessl2016drama}.
To this end, \codename enables users to test how vulnerable their computers are to the rowhammer problem and also help evaluate the impact of existing rowhammer attacks~\cite{xiao2016one,cheng2019cattmew,pessl2016drama, gruss2017another, aweke2016anvil} and rowhammer DRAM PUFs~\cite{schaller2017intrinsic,zeitouni2018s}.

In summary, we make the following key contributions:
\begin{itemize}
	\item We present a generic knowledge-assisted tool to reverse-engineer DRAM address mappings on both DDR3 and DDR4 chips for different CPU microarchitectures.
	\item Compared to existing works, our proposed algorithm makes full use of system DRAM information, which makes itself efficient in generating a deterministic DRAM address mapping. 
	\item We perform rowhammer tests based on \codename and have induced a significantly higher number of bit flips than previous works, justifying the correctness of our uncovered DRAM address mappings. 
\end{itemize}

\eat{
\mypara{Organization}
The rest of the paper is structured as follows.
In Section~\ref{sec:bkgd}, we introduce the DRAM organization and rowhammer vulnerability as well as related work.
In Section~\ref{sec:overview}, we describe our reverse-engineering algorithm in detail. 
Section~\ref{sec:eva} evaluates the algorithm on different CPU microarchitectures on different DRAMs including DDR3 and DDR4 and presents a detailed comparison of existing rowhammer test tools. 
At last, we conclude this paper in Section~\ref{sec:conclusion}.
}
\section{Background and Related Work}\label{sec:bkgd}
\eat{
先验知识放在overview，包括哪些部分的知识。
discussion latency测量部分，比已有算法要准，
interleaved mode 有些column bits和bank bits
supports DDR3 and DDR4
limitation and advantage
a hybrid solution 
decode-dimm system knowledge 
}

In this section, we first describe the modern DRAM organization and then briefly introduce the rowhammer vulnerability.
 
\subsection{Dynamic Random-Access Memory}
Dynamic Random-Access Memory (DRAM) modules are produced in the form of Dual Inline Memory Module (DIMM). A DIMM is directly connected to the CPU's memory controller through one of two channels. A DIMM consists of one or two ranks, corresponding to its one or two sides. Each rank is further decomposed of multiple banks. A bank is structured as arrays of cells with rows and columns. 
When a memory access to a desired bank occurs, this ``opens'' a specified row by transferring all data in the row to the bank's row buffer and a specified column from the row buffer will be accessed. As such, subsequent access to the same row will be served by the row buffer, while opening another row will flush the row buffer. 

\mypara{DRAM Refresh}
The charge in the DRAM cell is not persistent and will drain over time due to various charge leakage reasons~\cite{kim2014flipping}. To prevent data loss, a periodic re-charge or refresh is required for all cells. DRAM specification specifies that the DRAM refresh interval is typically 32 or 64 \emph{ms}, during which all cells within a rank will be refreshed. 

\mypara{DRAM Address Mapping}
The CPU's memory controller decides how physical-address bits are  mapped to a DRAM address. 
A DRAM address refers to a 3-tuple of \emph{bank, {row}, {column}} (DIMM, channel, and rank are included into the \emph{bank} tuple). As this mapping is not publicly documented on a major processor platform, i.e., Intel, Seaborn et al.~\cite{seaborndram} observed that only different rows within the same bank can induce rowhammer bit flips. Based on this observation, they made an educated guess on the DRAM address mapping of an Intel Sandy Bridge CPU. Both DRAMA~\cite{pessl2016drama} and Xiao et al.~\cite{xiao2016one} relied on a timing channel~~\cite{moscibroda2007memory} to uncover the mapping. 


\subsection{The Rowhammer Vulnerability}
DRAM rows are vulnerable to persistent charge leakage induced by adjacent rows. Specifically, frequently opening (also known as rowhammer) one row within the DRAM refresh interval can cause bit flips to the neighboring rows. To trigger expected rowhammer bit flips from a modern CPU, an adversary has to clear the CPU caches and the row buffer, and gain the knowledge of how DRAM is accessed by the CPU. \emph{Firstly}, all-level CPU cache must be flushed in order to hammer rows and this can be done explicitly by an unprivileged instruction (e.g., \texttt{clflush}) on x86 or implicitly by eviction sets of physical memory lines~\cite{aweke2016anvil, rowhammerjs, genkin2018drive}.
\emph{Secondly}, bypassing the row buffer is also a necessity. Intuitively, hammering (i.e., frequently accessing) two different rows within the same bank in an alternate manner can bypass the row buffer. If two rows happen to be one row apart, such technique is called \emph{double-sided} rowhammer. If not, then it is called \emph{single-sided} rowhammer. Alternatively, \emph{one-location} rowhammer~\cite{gruss2017another} forces the memory controller to clear the row buffer and thus only needs to hammer one row.
\emph{Lastly}, to map a virtual address to a DRAM address, Memory Management Unit (MMU) will translate the virtual address to a physical address, which is mapped to a DRAM address by the memory controller. The virtual to physical mapping can be addressed by either accessing \texttt{pagemap} or forcing \emph{hugepage} allocation. Clearly, a correct DRAM address mapping helps induce more rowhammer bit flips and thus affects existing rowhammer attacks~\cite{xiao2016one,cheng2019cattmew,pessl2016drama, gruss2017another, aweke2016anvil} and rowhammer DRAM PUFs~\cite{schaller2017intrinsic} as well as its attack~\cite{zeitouni2018s}.

\section{Overview}\label{sec:overview}

\begin{figure}[htbp]
\centerline{\includegraphics[width=\columnwidth, height=3.6in]{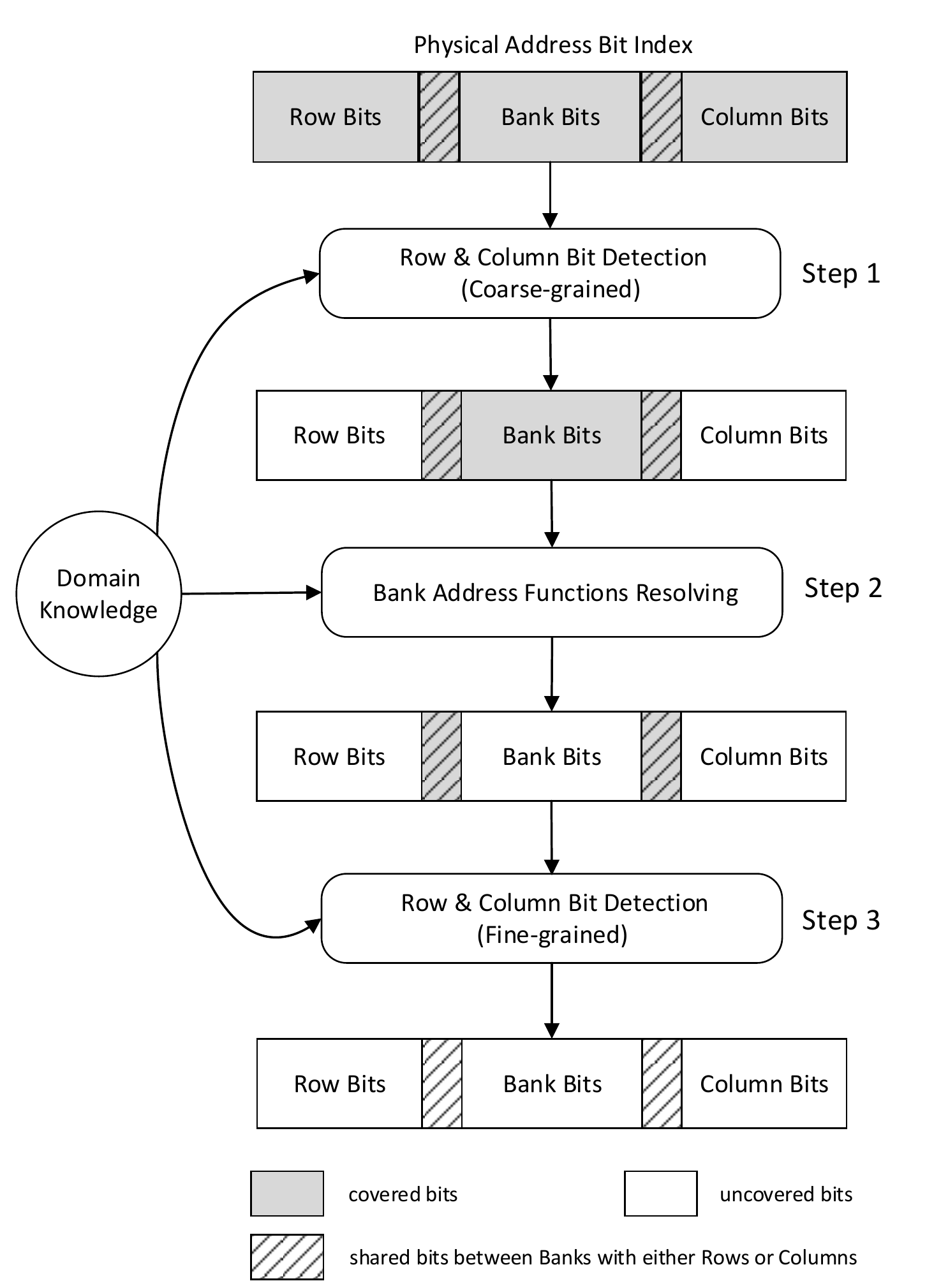}}
\caption{\codename Workflow.}
\label{fig:dramadig_arch}
\end{figure}

In this section, we describe a knowledge-assisted approach, named \codename,  to reverse engineer DRAM address mappings. As shown in Figure~\ref{fig:dramadig_arch}, \codename consists of three main steps. In \emph{Step 1}, we perform row and column bits detection and produce a coarse-grained results; that is, most row and column bits are uncovered while bits in grey boxes are still covered. In  \emph{Step 2}, we carefully select physical addresses that only differ in bits shown in grey boxes, then partition those addresses into different piles (addresses in each pile share the same bank), and identify bank address functions that can work for all piles. In \emph{Step 3}, we perform a fine-grained analysis on the resolved bank address functions to detect row or column bits that also play a role in the bank address functions as shared bits (see lined boxes in Figure~\ref{fig:dramadig_arch}). Note that each step needs assistance from specific domain knowledge.

In the following sections, we first introduce the domain knowledge and a timing primitive, and then discuss the major three steps of \codename in details.


\subsection{Domain Knowledge}\label{sec:domain}
We categorize the domain knowledge into three groups. 
\begin{itemize}
    \item \emph{Specifications}. We refer to DDR3 and DDR4 specifications~\cite{DDR3, DDR4} to acquire  physical-address bit numbers that index banks, rows and columns on specific DRAM chips.
    \item \emph{System Information}. This includes the total number of banks, physical memory size, and whether DRAM chips support ECC protection. This information can be obtained from the output of system commands such as \emph{decode-dimms} and \emph{dmidecode}.
    \item \emph{Empirical observations}. We have considered two aspects. 
    Firstly, a bank address function on Intel microarchitecture is a tuple of multiple physical address bits, which are XORed to output a single bit.
    Secondly, since Ivy Bridge, the lowest bit of a bank address function that owns the most number of bits is not the column bit.
        
\end{itemize}

\subsection{A Timing Primitive}
We resort to a timing channel~\cite{moscibroda2007memory} to reverse-engineer the DRAM mappings. Specifically, this timing channel is caused by the row-buffer conflicts within the same DRAM bank. As mentioned in Section~\ref{sec:bkgd}, each bank has a row buffer that caches the last accessed row. If a pair of addresses reside in two different rows of the bank and they are accessed alternately, the row buffer will be repeatedly reloaded and cleared. This causes the so-called row-buffer conflicts. Clearly, row buffer conflicts can lead to higher latency in accessing the two addresses than the case that they lie either within the same row or in different banks. As such, we can distinguish whether two addresses are in different rows within the same bank. 

\subsection{Row \& Column Bit Detection (Coarse-Grained)}\label{sec:coarse_grained_bit_detection}
We firstly partition the physical address bits into row, column and bank bits at a coarse-grained level. 
We use the same approach as the work~\cite{xiao2016one}. Specifically, for row bits, we measure access latency for two physical addresses that only with one bit different. If the latency is high, that means those addresses reside within the same bank but different row (SBDR), and that bit is the only different bit for the two addresses, so it is a row bit. 

For column bits, we select two physical addresses with only two bits being different, one being any detected row bit and the other being one non-row bit. If we get high latency for those two addresses, the non-row bit is a column bit. This because the two addresses are in the same bank, so neither of those bits is a bank bit, and then that non-row bit is a column bit that determine which column a physical address is mapped to. Finally, We consider the bits left are bank bits. 

Note that the bits detection results in this step are coarse-grained. The row and column bits detected so far are only responsible for addressing rows and columns. There could be other row bits or column bits that also take responsibility for addressing banks, and they are not detected in this step. We will discuss how to identify them in section~\ref{sec:fine_grained_bit_detection}.

\subsection{Bank Address Function Resolving}
We uncover the bank address functions based on the above coarse-grained detection results. It consists of three phases. First, we calculate a number of specific address ranges based on the bank bits, 
and then select physical addresses within those ranges. Second, we partition the selected addresses into \#bank number of piles, of which \#bank can be known from the \emph{System Information} in Section~\ref{sec:domain}. At last, we investigate those address piles and detect the bank address functions. We discuss the three phases in detail as follows.

\mypara{Physical-Address Selection}
The main idea is to only select the physical addresses within specific ranges. Such ranges reflect all possible values of all bank bits, so the selected addresses contain all the bank address functions. Besides, as the number of bank bits is determined through coarse-grained detection, the number of selected addresses can be determined. Algorithm~\ref{algo_phys_select} presents how we conduct physical-address selection.


\begin{algorithm}
\small
\caption{Physical-Address Selection} 
\label{algo_phys_select}
\begin{algorithmic}[1]
\Require $phys\_pages$: allocated memory pages; $B$: possible bank bits;
\Ensure $phys\_pool$
\State $b\_min, b\_max = find\_min\_max(B)$
\State $range\_mask = (1<<(b\_max+1)) - (1{<}{<}b\_min)$
\State $miss\_mask = 0$
\For {$b \in [b\_min, b\_max]$ \&\& $b \notin B$}
    \State $miss\_mask {+}{=} (1 << b)$
\EndFor
\For {$p \in phys\_pages$}
    \If {$(p {\&} range\_mask) == range\_mask$}
        \State $P\_start = p - range\_mask$
        \State $P\_end = p + PAGE\_SIZE$
        \If {!page\_miss($phys\_pages$, $P\_start$, $P\_end$)}
                \State break
        \EndIf
    \EndIf
\EndFor
\State $phys\_pool = \{\}$
\For {$p = P\_start$; $p < P\_end$; $p += (1<<b\_min)$}
    \State $p'$ = $p$ $|$ $miss\_mask$
    \If {$page(p') \in phys\_pages$}
        \State $phys\_pool.add(p')$
    \EndIf
\EndFor
\State \Return $phys\_pool$

\end{algorithmic}
\end{algorithm}

The algorithm first calculates a $range\_mask$, which indicates bank bits positions, and selects the physical pages that cover that range. (line 7-15). We don't want miss any bank bits and thus we require that the selected physical pages are consecutive, and if there are some pages missed in $phys\_pages$, we try again. The last found physical page range is presented by $[P\_start, P\_end]$.


Note that we use $b\_min$ and $b\_max$ to calculate range mask, but not all the bits in $[b\_min, b\_max]$ will be used in the bank address functions. We use ``$miss\_mask$" to represent those bits that have nothing to do with the address functions such that we set the bits to 1. Next, for every investigated address $p$ with $[P\_start, P\_end]$, we make it first mask against $miss\_mask$ and then check whether the new address $p'$ is valid and if so, we add it to $phys\_pool$, which holds final selected addresses. With ``$miss\_mask$", it enables us to only focus on the reasonable number of addresses that actually matter the address functions. 



\mypara{Physical Addresses Partition}
We apply Algorithm~\ref{algo_phys_partition} to partition the selected addresses into \#bank numbers of piles.

First, we randomly select one address $p$ from $phys\_pool$ and measure the latency with every other address. If the latency is high, it means we have one address that is SBDR with $p$. We put it into $piles[p]$ which stores addresses from $phys\_pool$ that are SBDR with $p$ (line 4-9). By doing so, we can collect all addresses from $phys\_pool$ that are SBDR with $p$ and record them in $piles[p]$. Next we verify whether the number of addresses in $piles[p]$ is within a valid range. If so, we consider this round of partition is successful, and then remove all the addresses in $piles[p]$ from $phys\_pool$ and conduct next round of partition (line 10-12). Finally, it stops when enough addresses have been partitioned (line 13-15).

\begin{algorithm}
\small
\caption{Physical-Address Partition}
\label{algo_phys_partition}
\begin{algorithmic}[1]
\Require: $phys\_pool$: selected physical addresses
\Ensure: $piles$: a map $<k, v>$, of which $k$ is an address and $v$ is the addresses that are SBDR with $k$. 
\State {$pool\_sz$ = $phys\_pool.size()$}
\State $pile\_sz$ = $pool\_sz/ \#bank$
\While {True}
    \State {randomly select p from $phys\_pool$}
    \For {$p' \in phys\_pool - \{p\}$}
        \If {$latency(p, p') == high$}
            \State $piles[p].add(p')$
        \EndIf
    \EndFor
    \If {$1-\delta \le piles[p].size() / pile\_sz \le 1+\delta$}
        \State $phys\_pool = phys\_pool - \{piles[p]\} - \{p\}$
    \EndIf
    \If {$phys\_pool.size() > per\_threshold * pool\_sz$}
        \State break
    \EndIf
\EndWhile
\State \Return $piles$
\end{algorithmic}
\end{algorithm}

Ideally, all the addresses in $phys\_pool$ will be partitioned into \#bank number of piles with each pile having the same number of addresses. However, in practice the partition may be influenced by noises introduced by incorrect results of latency measurement, so it is possible that not all of piles have the same number of addresses, and also there may be some addresses that not partitioned into any pile. That's why we introduce $\delta$ and $per\_threshold$ and they can be adjusted in practice. Empirically, we set $\delta$ to 0.2 and $per\_threshold$ to 85\% and then the addresses can be successfully partitioned into \#bank number of piles. 

\mypara{Bank Address Function Detection}
We utilize Algorithm~\ref{algo_bank_addr_func} to present how to detect bank address functions based on the address piles obtained.

\begin{algorithm}
\small
\caption{Bank Address Function Detection.}
\label{algo_bank_addr_func}
\begin{algorithmic}[1]
\Require: $piles$: a map $<k, v>$, of which $k$ is an address and $v$ is the addresses that have SBDR with $k$; $B$: bank bits
\Ensure:  $bank\_funcs$: the set that stores bank address functions
\State {$xor\_masks$ = gen\_xor\_masks($B$)}
\State {$bank\_funcs$ = \{\}}
\For {$pile \in piles$}
    \State {$func\_set$ = \{\}}
    \For {$mask \in xor\_masks$}
        \If {apply\_xor\_mask\_to\_pile($mask$, $pile$)}
            \State {$func\_set$.insert($mask$)}
        \EndIf
    \EndFor
    \State {$bank\_funcs$ = $bank\_funcs$ $\cap$ $func\_set$}
\EndFor
\State {prioritize($bank\_funcs$)}
\State {remove\_redundant($bank\_funcs$)} 
\State {check\_numbering($bank\_funcs$, $piles$)} 
\State \Return $bank\_funcs$
\end{algorithmic}
\end{algorithm}

\begin{table*}
\renewcommand\arraystretch{1.3}
\centering
\footnotesize
\begin{tabular}{ccccccc}
\hline
{\textbf{Machine}} & \multirow{2}{*}{\textbf{Microarch.}} & \multicolumn{2}{c}{\textbf{DRAM}}  & \multirow{2}{*}{\textbf{Bank Address Functions}} & \multirow{2}{*}{\textbf{Row Bits}} & \multirow{2}{*}{\textbf{Column Bits}} \\ \cline{3-4}
\textbf{No.} & & \textbf{Type, Size} & \textbf{Config.} &  &  & \\ 
\hline
\multirow{2}{*}{No.1} & {Sandy Bridge}  & \multirow{2}{*}{DDR3, 8GiB} & \multirow{2}{*}{2, 1, 1, 8} & \multirow{2}{*}{(6), (14, 17), (15, 18), (16, 19)} & \multirow{2}{*}{17$\sim$32} & \multirow{2}{*}{0$\sim$5, 7$\sim$13} \\
& i5-2400 &  &  &  &  & \\ 
\hline
\multirow{2}{*}{No.2} & {Ivy Bridge} & \multirow{2}{*}{DDR3, 8GiB} & \multirow{2}{*}{2, 1, 2, 8} & \multirow{2}{*}{(14, 18), (15, 19), (16, 20), (17, 21), (7, 8, 9, 12, 13, 18, 19 )}  & \multirow{2}{*}{18$\sim$32} & \multirow{2}{*}{0$\sim$6, 8$\sim$13} \\ 
&i5-3230M &  &  &  &  & \\
\hline
\multirow{2}{*}{No.3}&Ivy Bridge & \multirow{2}{*}{DDR3, 4GiB} & \multirow{2}{*}{1, 1, 2, 8} & \multirow{2}{*}{(13, 17), (14, 18), (15, 19), (16, 20)} & \multirow{2}{*}{17$\sim$31} & \multirow{2}{*}{0$\sim$12} \\
&i5-3230M &  &  &  &  & \\ 
\hline
\multirow{2}{*}{No.4}&Haswell & \multirow{2}{*}{DDR3, 4GiB} & \multirow{2}{*}{1, 1, 1, 8} & \multirow{2}{*}{(13, 16), (14, 17), (15, 18)} & \multirow{2}{*}{16$\sim$31} & \multirow{2}{*}{0$\sim$12} \\
&i5-4210U &  &  &  &  & \\ 
\hline
\multirow{2}{*}{No.5}&Haswell & \multirow{2}{*}{DDR3, 16GiB} & \multirow{2}{*}{2, 1, 2, 8} & \multirow{2}{*}{(14, 18), (15, 19), (16, 20), (17, 21), (7, 8, 9, 12, 13, 18, 19 )}  & \multirow{2}{*}{18$\sim$32} & \multirow{2}{*}{0$\sim$6, 8$\sim$13} \\
&i7-4790 &  &  &  &  & \\ 
\hline
\multirow{2}{*}{No.6}&Skylake & \multirow{2}{*}{DDR4, 16GiB} & \multirow{2}{*}{2, 1, 2, 16} & \multirow{2}{*}{(7, 14), (15, 19), (16, 20), (17, 21), (18, 22), (8, 9, 12, 13, 18, 19)}  & \multirow{2}{*}{19$\sim$33} & \multirow{2}{*}{0$\sim$7, 9$\sim$13} \\
&i5-6600 &  &  &  &  & \\ 
\hline
\multirow{2}{*}{No.7}&Skylake & \multirow{2}{*}{DDR4, 4GiB} & \multirow{2}{*}{1, 1, 1, 8} & \multirow{2}{*}{(6, 13), (14, 16), (15, 17)} & \multirow{2}{*}{16$\sim$31} & \multirow{2}{*}{0$\sim$12} \\
&i5-6200U &  &  &  &  & \\ 
\hline
\multirow{2}{*}{No.8}&Coffee Lake & \multirow{2}{*}{DDR4, 8GiB} & \multirow{2}{*}{1, 1, 1, 16} & \multirow{2}{*}{(6 13), (14 17), (15 18), (16, 19)}  & \multirow{2}{*}{17$\sim$32} & \multirow{2}{*}{0$\sim$12} \\
&i5-9400 &  &  &  &  & \\ 
\hline
\multirow{2}{*}{No.9}&Coffe Lake & \multirow{2}{*}{DDR4, 16GiB} & \multirow{2}{*}{2, 1, 2, 16} & \multirow{2}{*}{(7, 14), (15, 19), (16, 20), (17, 21), (18, 22), (8, 9, 12, 13, 18, 19)}  & \multirow{2}{*}{19$\sim$33} & \multirow{2}{*}{0$\sim$7, 9$\sim$13}\\
&i5-9400 &  &  &  &  & \\ 
\hline

\end{tabular}
\caption{Reverse-Engineered DRAM Mappings on 9 different machine settings. (The \textbf{Config.} column presents a specific DRAM configuration in a quadruple: (channel (\#), DIMMs (\#) per channel, ranks (\#) per DIMM, banks (\#) per rank).)}
\label{tab:dram_rev_engineer_result}
\end{table*}

According to \emph{Empirical Observation} in Section~\ref{sec:domain}, bank address functions take some bank bits as input and output $XORed$ values from those bank bits. Since we have grouped \#bank number of piles and the addresses in each pile map to the same bank, we look into each pile and try all combinations of bank bits and apply each of them to the addresses in the pile. We look into the combination starting from one bit to the number of bank bits (line 1). If a combination of bank bits has the same XORed result for all the addresses in the pile, we consider it as a possible bank address function. After investigating all the piles, we can have all possible bank address functions (line 3-11). 

However, some address functions are just linear combinations of the others so they are not the actual address functions and need to be removed. We consider the functions that have fewer bits have higher priority and remove the lower one if it is the linear combinations of higher priority functions. For instance, if (14, 18), (15, 19) and (14, 15, 18, 19) are 3 bank address functions. We consider the previous two have higher priority than the third, because the third is the linear combination of the previous two and we consider it as redundant (line 12-13).

Apart from that, the number of bank address functions should be $log_2(\#bank)$. There may be more than that number of functions after removing the redundant, and they are not actual address functions. So we test every combination of $log_2(\#bank)$ number of functions and consider them as bank address functions, and use them to number the address piles. The actual bank address functions should be able to count those piles from $0$ to $\#bank-1$ (line 14). 






\subsection{Row \& Column Bit Detection (Fine-Grained)}\label{sec:fine_grained_bit_detection}

As discussed in section~\ref{sec:coarse_grained_bit_detection}, we need to determine the row bits and column bits that are also bank bits. From the \emph{Specifications} in Section~\ref{sec:domain}, we can know the exact number of row and column bits for a specific DRAM chip, and since we have detected some row bits and column bits, we are able to determine how many row bits and column bits left to be uncovered. 

For the remaining covered row bits, we start to investigate the bank address functions that consist of two bits. We select two physical address with only those two bits different and measure their latency. The two addresses actually map to the same bank. If the latency is actually high, it means either one bit is a row bit. We consider the higher one as the row bits as discussed in ~\cite{seaborndram,xiao2016one}. 
If there are also row bits still covered after investigating two-bit address functions, we proceed to investigate address the bank functions that have more bits. In practice, we have not seen any case that needs to investigate the address functions that have three or more bits. 


For the remaining covered column bits, we first check the number of remaining column bits that need uncovered, and then identify those bits that have not been identified as column bits in coarse-grained detection, denoted as $C$. Next, according to the \emph{Empirical Observation} in Section~\ref{sec:domain}, we know the lowest bit (denoted as $l$) of the function occupying the most number of bits is not a column bit. So we investigate the bits $\{C-l\}$, by the order from low to high, and consider the first requested number of bits as column bits.

\eat{
\subsection{Distinguishing our algorithm}
Compared to DRAMA~\cite{pessl2016drama} and Xiao et al.~\cite{xiao2016one}, our approach is more robust and thus able to obtain more stable and complete result. DRAMA~\cite{pessl2016drama} blindly selected addresses and make partitions so it is more likely to produce incomplete or even incorrect DRAM mapping results. Besides, it does not guarantee that every test will produce the same results. Besides, Xiao et al.~\cite{xiao2016one} proposed a graph based approach but it only takes consideration of limited number of cases, so it is less likely to uncover DRAM mappings that never seen. By comparison, our method carefully selects specific range of physical addresses and applies completed searching policy to uncover address functions, so it is able to produce more stable and complete result.
}



\section{Evaluation}\label{sec:eva}
In this section, we first present DRAM address mappings uncovered by \codename, and then evaluate its performance overhead. 
At last, we utilize the uncovered mappings on 3 machines to perform double-sided rowhammer tests.

\subsection{Uncovered DRAM Address Mappings}\label{sec:dram mapping}
We test \codename on multiple Linux systems with different combinations of Intel microarchitectures and DRAM chips including DDR3 and DDR4. \codename has successfully uncovered DRAM mappings including row and column bits and bank address functions for all the test settings, as shown in Table~\ref{tab:dram_rev_engineer_result}.
From the table, we see that \codename has uncovered DRAM address mappings not only for mainstream CPU microarchitectures but also for a much newer CPU architecture (i.e., Coffee Lake) that previous works have never discussed. 

\eat{
\begin{table}[!htbp]
\caption{DRAM Settings for Experiments}
\label{tab:DRAM_settings}
\centering
\footnotesize
\begin{tabular}{cccccc}
\hline
\multirow{2}{*}{No.} & \multirow{2}{*}{CPU Family} & \multirow{2}{*}{CPU Name} & \multirow{2}{*}{Config.} & \multirow{2}{*}{Mem Size} & \multirow{2}{*}{Type} \\
 &  &  & & \\ \hline
1 & Sandy Bridge & i5-2400 & 2, 1, 1, 8 & 8G & DDR3\\
2 & Ivy Bridge & i5-3230M & 2, 1, 2, 8 & 8G & DDR3\\ 
3 & Ivy Bridge & i5-3230M & 1, 1, 2, 8 & 4G & DDR3\\
4 & Haswell & i5-4120U & 1, 1, 1, 8 & 4G & DDR3\\
5 & Haswell & i5-4790 & 2, 1, 2, 8  & 16G & DDR3\\
6 & Skylake & i5-6600 & 2, 1, 2 ,16 & 16G & DDR4\\
7 & Skylake  & i5-6200U & 1, 1, 1, 8 & 4G & DDR4 \\
8 & Coffee Lake & i5-9400 & 1, 1, 1, 16 & 8G & DDR4\\
9 & Coffee Lake & i5-9400 & 2, 1, 2, 16 & 16G & DDR4\\
\hline
\end{tabular}
\end{table}
}

\mypara{Distinguish \codename}
When executing the code that Xiao el al.~\cite{xiao2016one} shared with us, we found that it could not work on more than half of the machine settings, that is, No.2 and No.6-9. Take the No. 6 machine setting as an instance, the running code was stuck after resolving (16, 20), (17, 21), (18, 22) as 3 of 6 bank address functions. As such, their approach is not generic.
%
For DRAMA~\cite{pessl2016drama}, we ran their code for numerous times and found that it generated different DRAM mappings most of the time. As DRAMA is the first generic reverse-engineering tool, we further compare \codename with it in terms of performance overhead and rowhammer tests in the following sections.

\subsection{Performance Overhead}
\begin{figure}[htpb]
\centerline{\includegraphics[width=\columnwidth, height=1.8in]{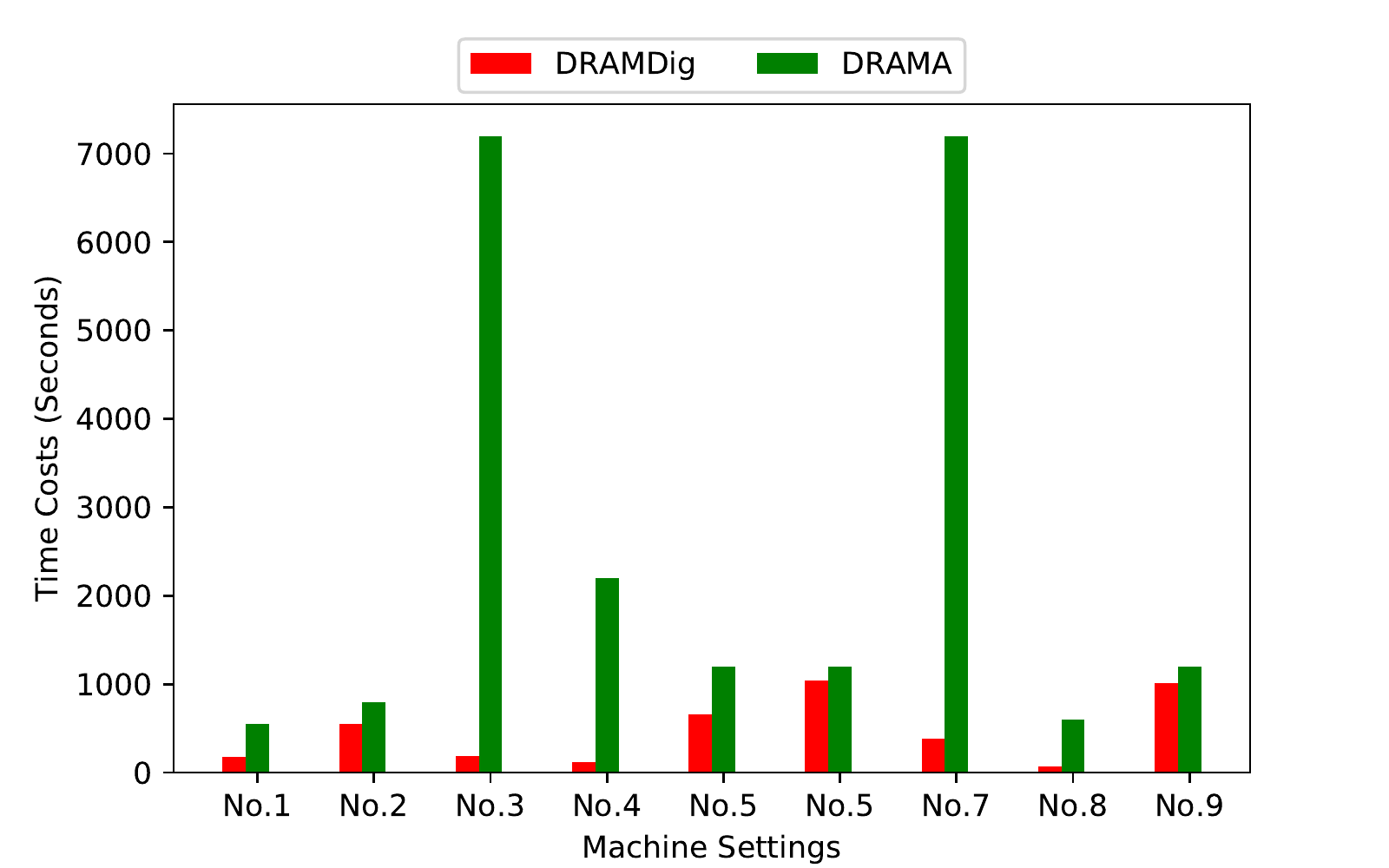}}
\caption{Time costs for \codename and DRAMA to uncover DRAM mappings on 9 machine settings. Clearly, \codename requires much less time than DRAMA on average.}
\label{fig:runtime_performance}
\end{figure}

\eat{We present the performance of \codename on DRAM mapping reverse engineering in Fig.~\ref{fig:runtime_performance}. As shown in the figure, \codename can produce DRAM mappings for different settings within 69 seconds to 17 minutes, about 7.8 minutes on average. Of all the stages of our approach, physical address partition took the very most proportion of the time. This part of time cost is mainly influenced by the number of selected physical addresses. The more addresses selected means the more latency measurements were needed for the partition. Specifically, for the test No.6 and No.9, \codename selected the most number of addresses (almost 16,000) while for the test No.8 it selected the least (about 4000).} 

We present the performance overhead required by \codename and DRAMA~\cite{pessl2016drama} respectively. As shown in Fig.~\ref{fig:runtime_performance}, \codename can be finished from 69 seconds to 17 minutes in all machines settings (only 7.8 minutes on average). In comparison, 
DRAMA spent from almost 500 seconds to 2 hours, indicating a much higher time cost than \codename. Particularly in No.3 and No.7 settings, it cost roughly two hours without producing any results before we killed it. 

For \codename, most of the time cost comes from the physical address partition, which is heavily affected by the number of selected physical addresses. The more selected addresses require more access latency measurements and thus the partition costs more time. Specifically, \codename in No.6 and No.9 settings selected the highest number of physical addresses (almost 16,000), making itself timing-consuming. In contrast, it selected the least number (about 4000) in No.8 setting, thus making itself time-saving. 

\eat{
\begin{figure}
\centerline{\includegraphics[width=\columnwidth, height=1.8in]{image/partition.pdf}}
\caption{Physical address partition comparison with DRAMA. The address number of each set partitioned by DRAMA varies greatly from each other, while each set partitioned by \codename has almost the uniform number of addresses.}
\label{fig:phy_addr_partition}
\end{figure}
}

\subsection{Rowhammer Tests}\label{sec:Rowhammer_Test}

A correct DRAM address mapping is critical in effectively inducing rowhammer bit flips and thus can justify the correctness of uncovered DRAM address mappings. Since both Seaborn's~\cite{seaborn2015exploiting} and Xiao's~\cite{xiao2016one} tools are not generic and limited to one or multiple machine settings, we compare our work with another generic tool, DRAMA~\cite{pessl2016drama}.

Specifically, we conducted double-sided rowhammer tests using the DRAM address mappings uncovered by \codename and DRAMA respectively on three machine settings. For each setting, we performed 5 rowhammer tests and each lasted 5 minutes. The results are induced bit flips as shown in Table~\ref{tab:rowhammer_test}. T1-T5 represent the 5 tests while the bit-flip numbers of each test are displayed in a way of \codename/DRAMA. From the \textbf{Total} column, we can clearly see that \codename caused significantly more bit flips than DRAMA while DRAMA even failed to induce any bit flips in some tests on No.2 and No.5. Also, the rowhammer test results justify the correctness of our uncovered DRAM address mappings.






\begin{table}[!htbp]
\centering
\resizebox{\columnwidth}{12mm} {
\begin{tabular}{ccccccc}
\hline
\multirow{1}{*}{\textbf{Machine}} & \multirow{1}{*}{\textbf{T1}} & 
\multirow{1}{*}{\textbf{T2}} & 
\multirow{1}{*}{\textbf{T3}} & 
\multirow{1}{*}{\textbf{T4}} & 
\multirow{1}{*}{\textbf{T5}} & 
\multirow{1}{*}{\textbf{Total}} \\
\multirow{1}{*}{\textbf{No.}} & \multicolumn{6}{c}{\multirow{1}{*}{\textbf{\codename/DRAMA}}} \\
\hline
\multirow{2}{*}{No.1} & \multirow{2}{*}{296/197} & \multirow{2}{*}{418/179} & \multirow{2}{*}{226/177}  & \multirow{2}{*}{627/259} & \multirow{2}{*}{484/286} & \multirow{2}{*}{2051/1098} \\
& & & & & & \\
\hline
\multirow{2}{*}{No.2} & \multirow{2}{*}{959/240} & \multirow{2}{*}{934/0} & \multirow{2}{*}{976/18} &
\multirow{2}{*}{1039/947} & \multirow{2}{*}{955/670} & \multirow{2}{*}{4863/1875} \\
& & & & & & \\
\hline
\multirow{2}{*}{No.5} & \multirow{2}{*}{12/7} & \multirow{2}{*}{12/0} & \multirow{2}{*}{12/0} & \multirow{2}{*}{10/0} & \multirow{2}{*}{11/0}  & \multirow{2}{*}{57/7} \\
& & & & & & \\
\hline
\end{tabular}
}
\caption{
Both \codename and DRAMA perform 5 tests on three machine settings and each test lasts 5 minutes. Clearly, \codename has induced significantly more bit flips.}
\label{tab:rowhammer_test}
\end{table}

\section{Conclusion}\label{sec:conclusion}

In this paper, we proposed a generic knowledge-assisted tool, \codename, which applied domain knowledge to efficiently and deterministically reverse-engineer DRAM address mappings. We evaluated \codename on a bunch of machines with different DRAM chips and CPU microarchitectures settings. The experiments showed that \codename was able to deterministically uncover DRAM address mappings on all the test machines with only 69 seconds in the best case and 17 minutes in the worst case. We also performed double-sided rowhammer tests with our uncovered DRAM address mappings and made comparison with DRAMA, and the results indicated that \codename could induce significantly more bit flips, which justified the correctness of our tool. As a result, we are going to open-source our tool to help users better understand DRAM address mapping and evaluate the impact of rowhammer attacks and rowhammer DRAM PUFs.

{
\footnotesize
\bibliographystyle{abbrv}
}
\bibliography{main}

\eat{
\section*{References}

Please number citations consecutively within brackets \cite{b1}. The 
sentence punctuation follows the bracket \cite{b2}. Refer simply to the reference 
number, as in \cite{b3}---do not use ``Ref. \cite{b3}'' or ``reference \cite{b3}'' except at 
the beginning of a sentence: ``Reference \cite{b3} was the first $\ldots$''

Number footnotes separately in superscripts. Place the actual footnote at 
the bottom of the column in which it was cited. Do not put footnotes in the 
abstract or reference list. Use letters for table footnotes.

Unless there are six authors or more give all authors' names; do not use 
``et al.''. Papers that have not been published, even if they have been 
submitted for publication, should be cited as ``unpublished'' \cite{b4}. Papers 
that have been accepted for publication should be cited as ``in press'' \cite{b5}. 
Capitalize only the first word in a paper title, except for proper nouns and 
element symbols.

For papers published in translation journals, please give the English 
citation first, followed by the original foreign-language citation \cite{b6}.

\vspace{12pt}
\color{red}
IEEE conference templates contain guidance text for composing and formatting conference papers. Please ensure that all template text is removed from your conference paper prior to submission to the conference. Failure to remove the template text from your paper may result in your paper not being published.
}
\end{document}